\documentclass[prl,twocolumn,superscriptaddress]{revtex4-1}

\usepackage[dvipdfmx]{graphicx}
\usepackage{type1cm}
\usepackage[usenames,svgnames]{xcolor}
\usepackage{color}
\usepackage{url}
\usepackage{natbib}
\usepackage[dvipdfmx]{hyperref}
\hypersetup{
     colorlinks=true,       		
     linkcolor=Navy,          	
     citecolor=Navy,            
     filecolor=Navy,      		
     urlcolor=Navy,           	
    runcolor=cyan,
 }

\usepackage{amsmath,amssymb,bm,amsthm}

\def\beq{\begin{equation}}
\def\eeq{\end{equation}}
\def\nbeq{\begin{equation*}}
\def\neeq{\end{equation*}}

\def\<{\langle}
\def\>{\rangle}

\newcommand{\sectionprl}[1]{{\par\it #1.---}}

\begin{document}
\title{Rigorous Bound on Energy Absorption and Generic Relaxation in Periodically Driven Quantum Systems}

\author{Takashi Mori}
\affiliation{
Department of Physics, Graduate School of Science,
University of Tokyo, Bunkyo-ku, Tokyo 113-0033, Japan
}
\author{Tomotaka Kuwahara}
\affiliation{
Department of Physics, Graduate School of Science,
University of Tokyo, Bunkyo-ku, Tokyo 113-0033, Japan
}
\affiliation{
WPI, Advanced Institute for Materials Research, Tohoku University, Sendai 980-8577, Japan
}
\author{Keiji Saito}
\affiliation{
Department of Physics, Keio University, Yokohama 223-8522, Japan
}

\begin{abstract}
We discuss the universal nature of relaxation in isolated many-body quantum systems subjected to global and strong periodic driving. Our rigorous Floquet analysis shows that the energy of the system remains almost constant up to an exponentially long time in frequency for arbitrary initial states and that an effective Hamiltonian obtained by a truncation of the Floquet-Magnus expansion is a quasi-conserved quantity in a long timescale. These two general properties lead to intriguing classification on the initial stage of relaxation, one of which is similar to the prethermalization phenomenon in nearly-integrable systems.
\end{abstract}
\maketitle

\sectionprl{Introduction}
In periodically driven many-body quantum systems, excited states as well as the ground state participate in the dynamics, and nontrivial macroscopic phenomena can appear. Recent years have witnessed remarkable experimental developments, such as the discoveries of the Higgs mode in the oscillating order parameter of the superconducting material under a Terahertz laser \cite{matsunaga2014light}, and the Floquet topological states in the periodically driven cold atom \cite{Aidelsburger2013,Atala2013,jotzu2014experimental,Aidelsburger2015}. Periodic driving in isolated quantum systems sometimes generates unexpected dynamical phenomena even if the instantaneous Hamiltonian at each time step is simple. To name only a few, dynamical localization~\cite{Dunlap-Kenkre1986,Grifoni_review1998,Kayanuma-Saito2008}, coherent destruction of tunneling~\cite{Grossmann1991,Grifoni_review1998,Kayanuma-Saito2008}, dynamical freezing~\cite{Das2010,Hegde2014}, and dynamical phase transitions~\cite{Prosen-Ilievski2011,Bastidas2012,Shirai2014a} are remarkable far-from-equilibrium phenomena that cannot be captured within linear-response analysis.  

On the other hand, as recently discussed in the context of {\em thermalization}, careful consideration is necessary on the true steady state in driven many-body systems~\cite{Russomanno2012,D'Alessio2013,Lazarides2014a,Lazarides2015,Ponte2015}.
Thermalization in isolated quantum systems has become one of critical subjects in modern physics \cite{Deutsch1991,Srednicki1994,Tasaki1998,Rigol2008,Polkovnikov_review2011,Sato2012}. The first study was made by von Neumann early in 1929~\cite{Neumann1929}, and now we are on the new stage by incorporating many concepts including quantum entanglement~\cite{Popescu2006} and experiments~\cite{Bloch_review2008}. In the case without driving fields, the notion of the eigenstate thermalization hypothesis (ETH) is a key idea~\cite{Neumann1929,Deutsch1991,Srednicki1994,Rigol2008} that states that each energy eigenstate is indistinguishable from the microcanonical ensemble with the same energy. As a generalization of ETH to periodically driven systems, the Floquet ETH was proposed, which states that all the Floquet eigenstates look the same and are indistinguishable from the infinite-temperature (i.e., completely random) state~\cite{D'Alessio2013,D'Alessio2014,Ponte2015,Lazarides2014b,Kim2014}. This leads to the conclusion that in general periodically driven many-body systems will eventually reach the steady state of infinite temperature, although several exceptions exist~\cite{Russomanno2012,Lazarides2014a,Lazarides2015,Ponte2015}.

The question that follows is on the time scale to reach the steady state. Recent experiments seem to urge us to clarify the general aspects of the time scale especially for the strong amplitude of global driving, where nontrivial transient dynamics is anticipated. We note that most nontrivial dynamical phenomena in driven systems are far-from-equilibrium effects that cannot be analyzed within linear-response analysis. Hence, in this paper, we for the first time aim to find the universal nature of the relaxation to the steady state under \textit{strong and global driving}. This direction is clearly crucial for a deeper understanding of thermalization and for analyzing the stability of transient dynamics in experiments. 

For this aim, we focus on the Floquet Hamiltonian $H_F$ which plays a central role in periodically driven systems:
\begin{eqnarray}
e^{-iH_F T} &\equiv & \mathcal{T}e^{-i\int_0^TdtH(t)} \, , \label{fldef}
\end{eqnarray}
where $H(t)$ is the Hamiltonian of the system, $\mathcal{T}$ is the time-ordering operator, and $T$ is the period of the driving ($\hbar=1$ throughout this paper). The Floquet Hamiltonian is an effective Hamiltonian that contains full information on the stroboscopic dynamics. The Floquet-Magnus (FM) expansion is a formal expression for the Floquet Hamiltonian: $H_F=\sum_{m=0}^{\infty} T^m \Omega_m$~\cite{Blanes_review2009,Bukov_review2015}. The explicit form of $\Omega_m$ is given in Eq.~(\ref{eq:FM}) below. However, it has recently been recognized that using full series expansion is problematic since it is not convergent in general. The convergence radius shrinks as the system size increases~\cite{Bukov_review2015}. Instead we here use the technique of truncation in the FM expansion, which was recently developed for describing the Floquet Hamiltonian for transient time scales \cite{Mori2015_Floquet,Kuwahara2016Floquet}: 
\begin{eqnarray}
e^{-iH_F^{(n)}T} &\simeq &  e^{-iH_FT}, ~~{\rm where}~~ H_F^{(n)}=\sum_{m=0}^nT^m\Omega_m \, . ~~ 
 \label{trfm}
\end{eqnarray}
Here, $H_F^{(n)}$ is the $n$th order truncated Floquet Hamiltonian. There are several studies which show that the time-evolution by the truncated Floquet Hamiltonian is reliable up to a certain long time $\tau$ for the driving with high frequency $\omega=2\pi/T$~\cite{Bukov_review2015}; $\tau\sim\omega^{1/2}$ for the Friedrichs model on the continuous space~\cite{Mori2015_Floquet} and $\tau\sim \exp[\mathcal{O}(\omega)]$ for lattice systems when driving is local~\cite{Kuwahara2016Floquet} or interactions are short-ranged~\cite{Kuwahara2016Floquet,Abanin_arXiv2015a,Abanin_arXiv2015b}. In this paper, we use the truncation technique for high frequency driving.

With this technique, two findings are mainly presented. We show as the first result that in the case of a high-frequency driving, the truncated Floquet Hamiltonian is a quasi-conserved quantity (a quantity that is almost conserved in a long timescale).
We also show as the second result that energy absorption rate per one site is bounded for an arbitrary amplitude of driving and for arbitrary initial states: 
\begin{eqnarray}
\dot{E}/N &\lesssim & (N_V/N)\exp[-\mathcal{O}(\omega/g)] \, , \label{firstresult}
\end{eqnarray}
where $E$ and $N$ are respectively the total energy and the number of lattice sites, and $g$ is the maximum energy per one site. The driving field is applied to $N_V$ sites. This provides a criterion on stability of transient quantum dynamics in experiments. These two findings lead to intriguing classification on the relaxation processes, one of which is similar to the prethermalization phenomenon seen in
non-driven nearly-integrable systems~\cite{Berges2004,Moeckel2008,gring2012relaxation,kollar2011generalized}, see Refs.~\cite{Bukov2015,Canovi2016}  for recent relevant numerical calculations.

\sectionprl{Setup and numerical example}
We consider a quantum spin system defined on a lattice with $N$ sites in arbitrary dimension,
whose Hamiltonian is written as 
\begin{eqnarray}
H(t)&=&H_0+V(t) \, . \label{hamil}
\end{eqnarray}
The driving field $V(t)$ is applied to $N_V (\le N)$ sites and satisfies the periodicity in time $V(t)=V(t+T)$ with zero average over the single period. 
We mainly focus on the regime of high-frequency $\omega=2\pi/T$. 
Each lattice site $i=1,2,\dots N$ has its own spin.
The basic assumption on the Hamiltonian is that it is expressed in the form of
\beq
H(t)=\sum_{X:|X|\leq k}h_X(t),
\label{eq:k-local}
\eeq
where $X=\{i_1,i_2,\dots,i_{|X|}\}$ is a set of the lattice sites with $|X|$ being the number of sites in $X$, and $h_X(t)$ is an operator acting on the sites in $X$.
In addition, we assume that the single site energy is bounded in the sense that
\beq
\text{for any site $i$, } \sum_{X:X\ni i}\|h_X(t)\|\leq g
\label{eq:g}
\eeq
with some fixed positive constant $g$, where $\|\cdot\|$ denotes the operator norm.

The form of Eq.~(\ref{eq:k-local}) means that the Hamiltonian contains at most $k$-body interactions.
For most physical applications, we can consider the case of $k=2$.
In the case of spin-(1/2) systems, the most general form of the Hamiltonian~(\ref{eq:k-local}) with $k=2$ is
\beq
H(t)=\sum_{i=1}^N\bm{B}_i(t)\cdot\bm{\sigma}_i
+\sum_{i<j}^N\sum_{\alpha,\gamma=x,y,z}J_{ij}^{\alpha\gamma}(t)\sigma_i^{\alpha}\sigma_j^{\gamma},
\label{eq:H_spin}
\eeq
where $\bm{\sigma}_i=(\sigma_i^x,\sigma_i^y,\sigma_i^z)$ is the Pauli matrix of $i$th spin, $\bm{B}_i(t)$ is the local magnetic field at $i$th site, and $J_{ij}^{\alpha\gamma}(t)$ denotes the interaction between $i$th and $j$th spins.
We can explicitly confirm that this Hamiltonian can be brought into the form of Eq.~(\ref{eq:k-local}) by putting $h_{\{i\}}=\bm{B}_i\cdot\bm{\sigma}_i$ and $h_{\{i,j\}}=\sum_{\alpha,\gamma=x,y,z}J_{ij}^{\alpha\gamma}\sigma_i^{\alpha}\sigma_j^{\gamma}$.

\begin{figure}[t]
\begin{center}
\includegraphics[width=8.0cm]{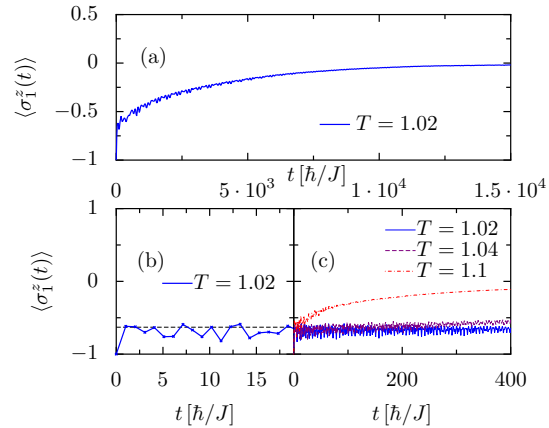}
\caption{(color online) Numerical demonstration of prethermalization-like phenomenon. (a): Relaxation in the long timescale, (b): initial relaxation, (c): transient time-evolution after the initial one. Parameters are $(J,B_x,B_z)=(1,0.9045, 0.8090)$ and $N=24$. The dotted line in (b) is the expectation value in the equilibrium state of $H_F^{(0)}$ at the inverse temperature $\beta=0.85$, which is determined from the expectation value of $H_F^{(0)}$ at $t=10$.}
\label{fig1}
\end{center}
\end{figure}
To make clear physical phenomena that we address, we show a numerical example with a toy model that has been used to show the Floquet ETH in \cite{Kim2014}. We consider dynamics over one cycle, taking $H(t)=H_z$ for the first half period and $H(t)=H_x$ for the second half period, where $H_z=\sum_{i=1}^N \left[ -J \sigma_i^z \sigma_{i+1}^z + B_z \sigma_i^z \right]$ with the periodic boundary condition and $H_x = B_x \sum_{i=1}^N \sigma_i^x$. We calculate the time-evolution of the $z$-component of the first spin $\sigma_1^z$ setting all spin-down state as the initial state. Floquet ETH implies that a steady state in the long-time limit is a random state, and hence, when it is satisfied, the expectation value of any local spin operator eventually reaches zero. In Fig.\ref{fig1}, the time-evolution for a sufficiently large system size is shown. Figure~\ref{fig1}(a), (b) and (c) are respectively time-evolution of $\sigma_1^z$ in the large timescale, the initial stage, and the transient timescale. Fig.\ref{fig1}(a) shows a vanishing expectation value that is a clear indication of the Floquet EHT. Crucial observation is that after the initial relaxation (Fig.\ref{fig1}(b)), the expectation value is almost constant for finite transient timescales, and the timescales depend on the period $T$ (Fig.\ref{fig1}(c)). This implies that the heating process is seemingly suppressed during this timescale. This is somewhat similar to the prethermalization phenomenon in nondriven nearly-integrable systems. In experimental situations, this transient time behavior is crucial, and hence we address the mechanism of the behavior and consider the period dependence on the timescale.

\sectionprl{Timescale of the heating process}
We use the FM expansion for analyzing the energy absorption and the relaxation process.
The FM expansion is the formal expansion of the Floquet Hamiltonian given by $H_F=\sum_{n=0}^{\infty}T^n\Omega_n$ with $\Omega_0=H_0$ and the $n$th order coefficient $\Omega_n$ for $n\geq 1$ being given by~\cite{Bialynicki-Biula1969}
\begin{widetext}
\begin{align}
\Omega_n&=&
\sum_{\sigma}
{ (-1)^{n-\theta[\sigma]}\theta[\sigma]!(n-\theta[\sigma])! \over 
i^n(n+1)^2n! T^{n+1} }
\int_0^Tdt_{n+1}
\dots\int_0^{t_2}dt_1
 [H(t_{\sigma(n+1)}),[H(t_{\sigma(n)}),\dots,[H(t_{\sigma(2)}),H(t_{\sigma(1)})]\dots]],
\label{eq:FM}
\end{align}
\end{widetext}
where $\sigma$ is a permutation and $\theta[\sigma]=\sum_{i=1}^n\theta(\sigma(i+1)-\sigma(i))$ with $\theta(\cdot)$ is the step function. 
It is believed that the FM expansion is divergent in many-body interacting systems~\cite{D'Alessio2014,Bukov_review2015,Ponte2015}. See the supplementary material for the numerical demonstration of the divergence \cite{supplement}. 
This divergence is not merely a mathematical phenomenon but is now thought to be an indication of heating process 
due to periodic driving~\cite{D'Alessio2014,Bukov_review2015,Ponte2015}.


We define the $n$th order truncation of the FM expansion as in Eq.~(\ref{trfm}) and 
show that for general spin systems the timescale of the heating is exponentially slow in frequency.
To this end, we start with an intuitive explanation on our analysis.
From Eq.~(\ref{eq:FM}), $\Omega_n$ has at most $(n+1)k$-spin effective interactions because of the multiple commutators in Eq.~(\ref{eq:FM}), which describes the collective flip of $(n+1)k$ spins.
Since the energy exchange between a quantum system and the external periodic field is quantized into integer multiples of $\omega$ and the energy of each spin is bounded by $g$, $N^*\sim\omega/g$ spins must flip cooperatively in order to absorb or emit the single ``energy quantum''.
Such a process is taken into account only in the terms higher than the $n_0$th order in the FM expansion with $n_0\sim N^*/k\sim\omega/(gk)$.
Indeed, each term of the FM expansion is rigorously bounded from above as
\beq
\|\Omega_n\|T^n\leq 2gN_V\frac{(2gkT)^nn!}{(n+1)^2} \, .
\label{eq:bound_Omega}
\eeq
This is given by estimating norms of the multiple commutators taking account that the Hamiltonian has at most $k$-body interaction and the energy per site is bounded by $g$ \cite{Kuwahara2016Floquet,supplement}. 
Equation (\ref{eq:bound_Omega}) shows that the FM expansion~(\ref{trfm}) looks convergent up to $n\leq n_0\sim \omega/(gk)$ and
\beq
\|H_F^{(n)}-H_F^{(n_0)}\|= N_V\mathcal{O}(T^{n+1}) \quad (n<n_0),
\label{eq:convergent}
\eeq
but grows rapidly for $n>n_0$.
Therefore, we can eliminate the heating effect most efficiently by truncating the FM expansion at $n=n_0$.
The timescale of the heating is thus evaluated by comparing the difference between the exact time evolution and the approximate time evolution under the $n_0$th order truncated Floquet Hamiltonian.
It is expected that higher-order terms (i.e. simultaneous flip of a large number of spins) would matter only in the later stage of the time evolution.

We now make the above argument mathematically rigorous. We can prove the following theorem:

\bigskip
\noindent
\textbf{Theorem.} \textit{The $n_0$th order truncated Floquet Hamiltonian $H_F^{(n_0)}$ is almost conserved up to an exponentially long time in frequency in the sense that}
\beq
\|H_F^{(n_0)}(t)-H_F^{(n_0)}\|\leq 16g^2k2^{-n_0}N_Vt,
\label{eq:conserved}
\eeq
\textit{where $t=mT$ with a positive integer $m$, $n_0=\lfloor 1/(8gkT)-1\rfloor$, and $H_F^{(n_0)}(t)=U^{\dagger}(t)H_F^{(n_0)}U(t)$ is the $n_0$th order truncated Floquet Hamiltonian at time $t$ in the Heisenberg picture with $U(t)=\mathcal{T}e^{-i\int_0^tdt'H(t')}$.}
\bigskip

\noindent
This is derived by evaluating the norm of the Dyson-expansion for the time-evolution operators in the left hand side taking into account that the Hamiltoanin is written as Eq.~(\ref{eq:k-local}) with Eq.~(\ref{eq:g}). See the supplementary material for more details on the derivation \cite{supplement}. Combined with Eq.~(\ref{eq:convergent}), this theorem leads to 
\beq
\|H_F^{(n)}(t)-H_F^{(n)}\|\leq 16g^2k2^{-n_0}N_Vt+N_V\mathcal{O}(T^{n+1})
\eeq
for any $n<n_0$.
In particular, by substituting $n=0$, we obtain
\beq
\frac{1}{N}\|H_0(t)-H_0\|\leq\frac{N_V}{N}\left[16g^2k2^{-n_0}t+\mathcal{O}(T)\right].
\label{eq:exp_slow_heating}
\eeq
It is noted that the term of $\mathcal{O}(T)$ in Eq.~(\ref{eq:exp_slow_heating}) is independent of $t$.
Thus, the energy density remains constant within a small fluctuation of $\mathcal{O}(T)$ until an exponentially long time in frequency. 
Equation~(\ref{eq:conserved}) provides {\em a lower bound} on the timescale during which $H_F^{(n_0)}$ can be an approximately conserved quantity. This quasi-conserving property lasts during the timescale larger than $\tau\sim 2^{n_0}\sim e^{\mathcal{O}(\omega)}$. Similarly, Eq.~(\ref{eq:exp_slow_heating}) implies that the lower bound of the timescale of heating is an exponentially long time in frequency. This is a main result (\ref{firstresult}). 
The exponentially long timescale of the energy relaxation was shown for short-range interacting spin systems in the linear-response regime in Ref.~\cite{Abanin2015}, but it should be emphasized that Eq.~(\ref{eq:exp_slow_heating}) has been obtained without assuming short-range interactions and the linear-response argument.
See Ref.~\cite{Bukov_arXiv2015} for a recent numerical result.

It is remarked that for local driving with $N_V\lesssim e^{\mathcal{O}(\omega)}$, a much stronger result was shown in Ref.~\cite{Kuwahara2016Floquet}, i.e., 
\beq
\|\mathcal{T}e^{-i\int_0^tH(s)ds}-e^{-iH_F^{(n_0)}t}\|\leq\exp[-\mathcal{O}(\omega)]t
\label{eq:local}
\eeq
for $t=mT$. This inequality implies that for any bounded operator that may be highly nonlocal, the FM truncated Hamiltonian gives the accurate time evolution up to an exponentially long time. 
In the case of global driving $N_V\propto N$, this strong inequality~(\ref{eq:local}) is not satisfied for sufficiently large systems, but even in this case, we can utilize the finite order truncation of the FM expansion to discuss the relaxation process as is argued below.

\sectionprl{Relaxation process}
Our rigorous result enables us to discuss possible scenarios on the initial stage of relaxation. According to the Floquet ETH~\cite{D'Alessio2014,Ponte2015,Lazarides2014b,Kim2014}, the steady state in the long-time limit induced by the Floquet Hamiltonian (\ref{fldef}) is a state of infinite temperature. Full FM series expansion in general diverges in large quantum systems and hence it is not useful~\cite{Bukov_review2015,Kuwahara2016Floquet}. 
However, the truncated Floquet Hamiltonian $H_F^{(n_0)}$ is a quasi-conserved quantity and plays a crucial role in the relaxation process.

We make a remark on the degree of nonlocality on the quasi-conserved quantity. The $n_0$th order truncated Floquet Hamiltonian has effective $(n_0+1)k$-body interactions, and hence the nonlocality looks large. However, for a high frequency driving, higher order contributions in the FM expansion are very small, since $T$ is small. The dominant contribution is in fact the original Hamiltonian $H_0$. Hence, nonlocality of the truncated Floquet Hamiltonian is not very strong. Eigenstates for the truncated Floquet Hamiltonian $H_F^{(n_0)}$ thus should satisfy the usual ETH, not the Floquet ETH. 
In addition, we should note that from Eq.~(\ref{eq:convergent}), $H_F^{(n)}\approx H_F^{(n_0)}$ for any $n<n_0$. Hence these truncated Floquet Hamiltonians are not independent but almost the same. Practically one can approximate the quasi-conserved quantity $H_F^{(n_0)}$ by $H_0 (=H_F^{(0)})$.

Taking account of those, we discuss a scenario on the initial stage of relaxation. Since the quasi-conserved quantity exists with long lifetime, the system relaxes to {\em a quasi-stationary state} characterized by the quasi-conserved quantity, which will be close to a state corresponding to the (micro)canonical ensemble $\rho_{\rm eq}^{(n_0)}$ of the effective Hamiltonian $H_F^{(n_0)}$ set by the initial state. Approximately one can use $\rho_{\rm eq}^{(0)}$ (the equilibrium ensemble of $H_0$) instead of $\rho_{\rm eq}^{(n_0)}$ because $H_F^{(n)}\approx H_F^{(n_0)}$ for any $n<n_0$.

The initial stage of relaxation can be classified into two cases, i.e., (i) the case where the relaxation to the quasi-stationary state is faster than the energy relaxation, and (ii) the case where both relaxation times are comparable. 
In the case (i), the system first reaches the quasi-stationary state, and then relaxes to the true steady state. This is highly related to the prethermalization phenomenon in the isolated nearly-integrable systems \cite{gring2012relaxation,kollar2011generalized}, where the system first relaxes to a quasi-stationary state corresponding to the generalized Gibbs ensemble and next relaxes to the true steady state. This is what we numerically observed in Fig.\ref{fig1} (b) and (c).
Remarkably, in Fig.~\ref{fig1} (b) and (c), $\<\sigma_1^z(t)\>\approx -0.65$ in the quasi-stationary state, which is close to ${\rm Tr}\rho_{\rm eq}^{(0)}\sigma_1^z$ (the dotted line in Fig.~\ref{fig1} (b)) at the inverse temperature $\beta=0.85$ that is determined from the expectation value of $H_F^{(0)}$ at $t=10$. This fact indicates that the quasi-stationary state is actually described by $\rho_{\rm eq}^{(0)}$ in this model. 

In the case (ii), on the other hand, the relaxation process towards the quasi-stationary state and that towards the true steady state are indistinguishable, and hence stable quasi-stationary behavior is not observed in the initial stage of relaxation.

Because the timescale of energy relaxation becomes longer exponentially as the frequency increases, we expect to find the case (i) for sufficiently high-frequencies unless there is some special reason such as conservation laws~\cite{Russomanno2012,Lazarides2014a}, strong quenched disorder~\cite{Lazarides2015,Ponte2015}, diverging timescale due to quantum criticality~\cite{Dziarmaga2005}, and so on.

Our analysis deals with general spin models, which makes clear why the heating is slow in a precise manner and leads us to the universal scenario of relaxation processes. 
However, in our evaluation, the single-site energy is overestimated and the effect of quantum interference is underestimated. Hence, the divergence of the FM expansion presumably begins at a higher order than our estimation $n_0\approx 1/(8gkT)$. We stress that our estimation on the timescale is {\em a rigorous lower bound} that can be exponentially large in frequency, and hence the actual timescale of the heating will be longer than our estimation~\footnote
{In cold atoms in an optical lattice, $g$ is typically about $100$ Hz, $k=2$, and then the condition $T<1/8gk$ implies $\omega\gtrsim 10$ kHz, which has been achieved in experiment [A. Zenesini, H. Lignier, D. Ciampini, O. Morsch, and E. Arimondo, \href{http://link.aps.org/doi/10.1103/PhysRevLett.102.100403}{Phys. Rev. Lett. {\bf 102}, 100403 (2009)}].
According to our estimation, the heating timescale in this case is about 1 msec, which is a typical timescale of cold-atom experiments.}. 
In order to obtain a quantitatively accurate estimate for a specific model, we will have to study the quantum dynamics of the given model numerically.

Related to the above remark, we emphasize that our result does not tell us about the true steady state.
It should be an infinite-temperature state if the Floquet ETH holds.
However, another possibility is not excluded; there might be an energy-localized phase~\cite{D'Alessio2013} with vanishing energy-absorption rate.
It is an open problem to understand the precise condition of the Floquet ETH.

\sectionprl{Summary}
In summary, we have considered the quantum dynamics of general driven spin systems that have at most $k$-body interactions and a bounded single-site energy $g$. We have rigorously shown the Theorem stating that the truncated Floquet Hamiltonian is a quasi-conserved quantity and the rate of energy absorption is exponentially small in frequency. This finding enables us to classify the initial stage of relaxation. 
It is emphasized that we need not assume short-range interactions in the Hamiltonian (\ref{eq:H_spin}).
For instance, $J_{ij}^{\alpha,\gamma}=\delta_{\alpha,\gamma}J/N$, which corresponds to the Heisenberg all-to-all couplings, satisfies the condition of Eq.~(\ref{eq:g}) with a fixed value of $g$ even in the thermodynamic limit.
Therefore, the result in this paper is applicable to most physically-relevant spin models.
However, as seen in Eq.~(\ref{eq:g}), our argument excludes bosonic systems.
We expect that our analysis will help to understand even for bosonic systems.

\begin{acknowledgments}
We are grateful to Naomichi Hatano and Hal Tasaki for critical reading of the manuscript.
T.M. was supported by the JSPS Core-to-Core Program ``Non-equilibrium dynamics
of soft matter and information'' and JSPS KAKENHI Grant No.~15K17718.
T.K. acknowledges the support from JSPS grant no.~2611111.
K.S. was supported by MEXT grant
no. 25103003.

\textit{Note added.} Recently, closely related results in a different approach have appeared~\cite{Abanin_arXiv2015a,Abanin_arXiv2015b}.
\end{acknowledgments}

\clearpage
\onecolumngrid
\begin{center}
{\large \bf Supplementary Material for  \protect \\ 
``Rigorous bound on energy absorption and generic relaxation in periodically driven quantum systems'' }\\
\vspace*{0.3cm}
Takashi Mori$^{1}$, Tomotaka Kuwahara$^{1}$ and Keiji Saito $^{2}$ \\
\vspace*{0.1cm}
$^{1}${\small \em Department of Physics, Graduate School of Science, University of Tokyo, Bunkyo-ku, Tokyo 113-0033, Japan} \\
$^{2}${\small \em Department of Physics, Keio University, Yokohama 223-8522, Japan} \\
\end{center}

\setcounter{equation}{0}
\setcounter{figure}{0}
\setcounter{table}{0}
\setcounter{page}{1}
\makeatletter
\renewcommand{\theequation}{S\arabic{equation}}
\renewcommand{\thefigure}{S\arabic{figure}}
\renewcommand{\bibnumfmt}[1]{[S#1]}
\section{Numerical demonstration of the FM expansion}

\begin{figure}[t]
\includegraphics[width=8.0cm]{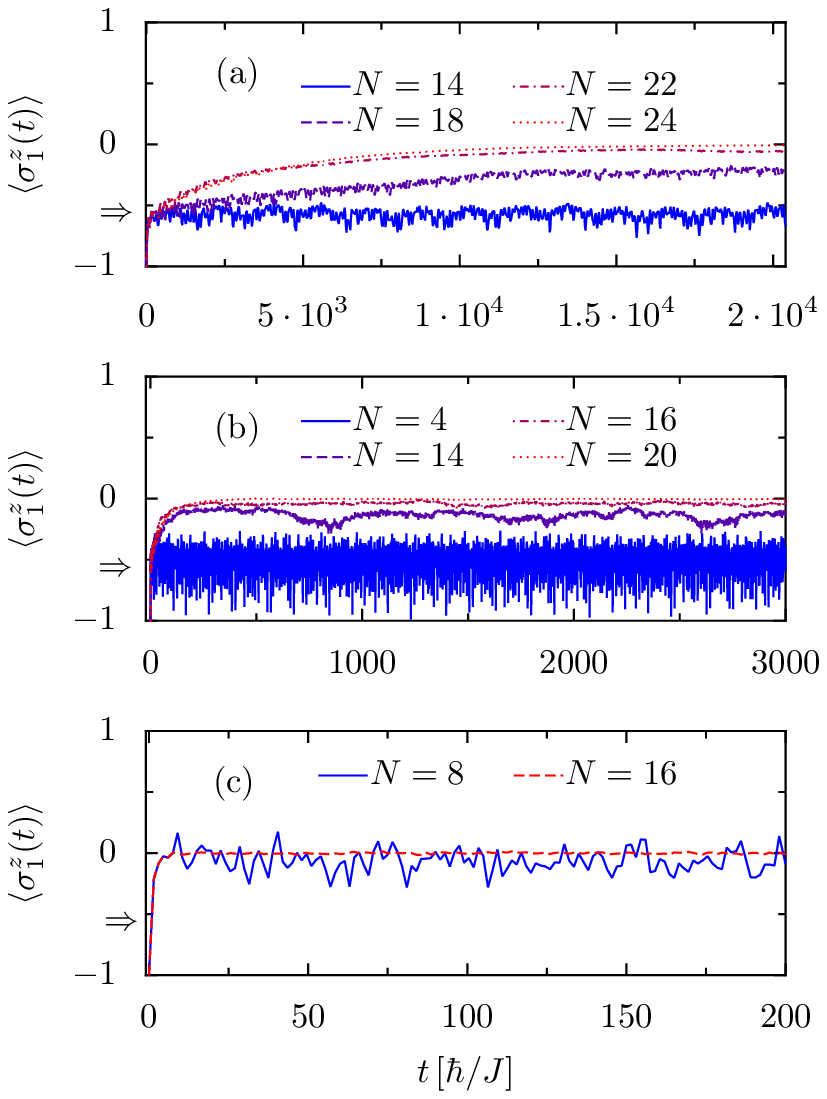}
\caption{(color online) Time-evolution of $\sigma_1^z$ for (a): $T=1.02$, (b): $T=1.2$, and (c) $T=1.5$.
The arrows indicate the equilibrium value with respect to $H_F^{(0)}$, i.e. ${\rm Tr}\rho_{\rm eq}^{(0)}\sigma_1^z$ with $\rho_{\rm eq}^{(0)}=e^{-\beta H_F^{(0)}}/{\rm Tr}e^{-\beta H_F^{(0)}}$.
The temperature is determined by the expectation value of $H_F^{(0)}$ in the initial state ($\beta=0.85$ in this case).}
\label{suppl1}
\end{figure}

We consider the same model that we discussed in the main text. Namely we take a toy model where the dynamics is driven by $H(t)=H_z$ for the first half period and $H(t)=H_x$ for the second half period. 
The time-evolution operator for the single period $U$ is given by 
\begin{eqnarray}
U &=& e^{-i H_x T/2 } e^{-i H_z T /2} \, .
\end{eqnarray}
We show the time-evolution of $\<\sigma_1^z(t)\>$ for several system sizes and periods in Fig.~\ref{suppl1}.
The initial state is chosen as the all spin-down state.
Each graph of Fig.~\ref{suppl1} shows that $\<\sigma_1^z(t)\>$ remains finite when the system size is small, while it eventually tends to zero when the system size is large.
This implies that a large system eventually reaches the state of infinite temperature, which is consistent with the Floquet ETH.
By comparing the three graphs of Fig.~\ref{suppl1}, we find that the timescale of the heating becomes longer as the period of the external field decreases.
This is also consistent with our result on the timescale of the heating.

The arrows in the figures indicate the expectation value of $\sigma_1^z$ in the equilibrium state of the 0th-order truncated Floquet Hamiltonian $H_F^{(0)}$, that is,  $\<\sigma_1^z\>_{\rm qss}={\rm Tr}\rho_{\rm eq}^{(0)}\sigma_1^z$ with $\rho_{\rm eq}^{(0)}=e^{-\beta H_F^{(0)}}/{\rm Tr}e^{-\beta H_F^{(0)}}$.
The temperature is determined from the expectation value of $H_F^{(0)}$ at $t=10$. In this case, $\beta=0.85$ and  $\<\sigma_1^z\>_{\rm qss}= -0.65$.
We can see that in the first stage of the relaxation, $\<\sigma_1^z(t)\>$ relaxes to $\<\sigma_1^z\>_{\rm qss}$, which indicates that a quasi-stationary state observed in this model is actually described by the equilibrium state of the truncated Floquet Hamiltonian.

\begin{figure}[t]
\includegraphics[width=12.0cm]{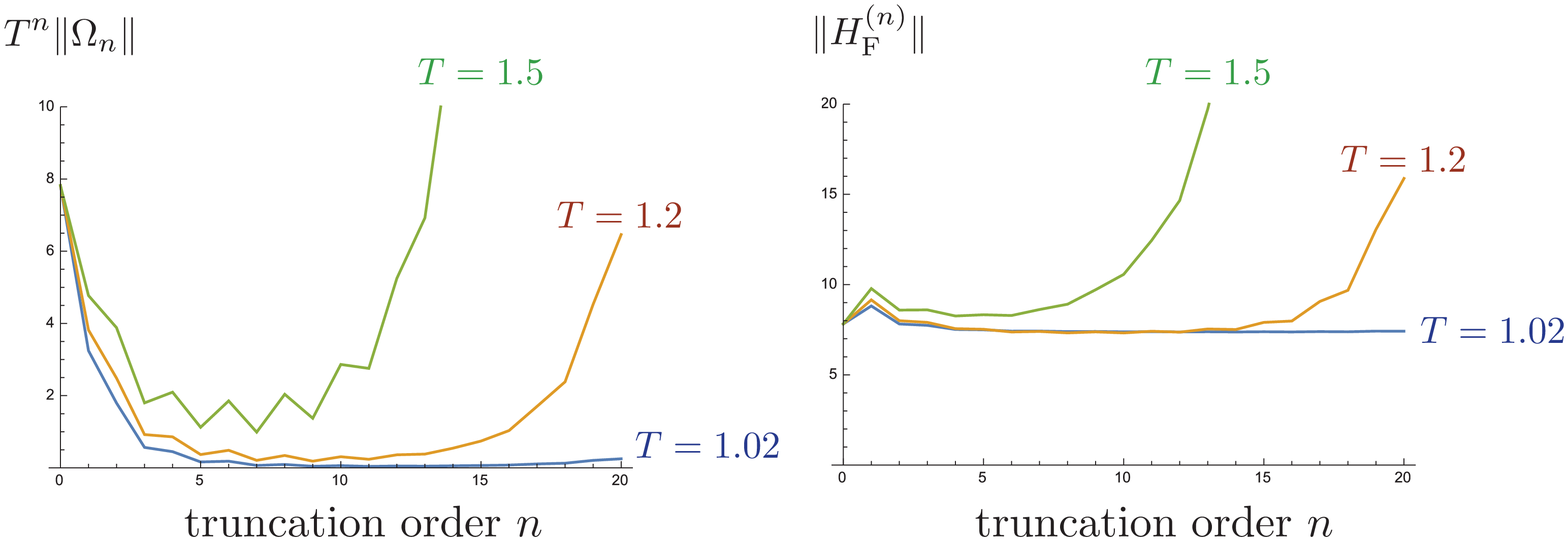}
\caption{(color online) The operator norm of $n$th order term of the FM expansion (left) and the $n$th order truncated Floquet Hamiltonian.}
\label{suppl2}
\end{figure}

We next demonstrate the numerical evidence that the FM-expansion can diverge in this model.
In Fig.~\ref{suppl2}, we show the operator norms of $\|H_F^{(n)}\|$ as a function of $n$.
It looks convergent up to some $n$ but it starts to diverge when $n$ exceeds this value.
As far as we computed, the divergence of the FM expansion occurs for $T\gtrsim 1.0$ when $N\leq 24$.

It is emphasized that the divergence begins at a higher order term than our estimation $n_0=\lfloor 1/(8gkT)-1\rfloor$.
Although the theorem presented in the main text becomes meaningless when $1/(8gkT)<1$ because $n_0$ becomes negative, the numerical simulations given here and in the main text show that exponentially slow heating is observed even in that case.
From this observation, we must conclude that although our rigorous analysis clearly shows that the timescale of heating is extremely slow in the high frequency regime for general spin systems, it does not give accurate quantitative details.
Numerical investigation is necessary in order to make a quantitatively accurate prediction for a given specific model.

\clearpage
\section{Proof of the Theorem}
\subsection{Preliminary}
Here, we provide the proof of the theorem. 
As a preliminary, we define the ``$k$-locality'' and the ``$g$-extensiveness'' as the properties of operators.
We say that an operator $A$ is $k_A$-local if it is decomposed as
\beq
A=\sum_{X:|X|\leq k_A}a_X,
\eeq
where $X$ is a set of lattice sites and $|X|$ is the number of the sites in $X$, see setup in the main text.
This operator $A$ is said to be $g_A$-extensive if
\beq
\sum_{X:X\ni i, |X|\leq k_A}\|a_X\|\leq g_A
\eeq
for all lattice sites $i=1,2,\dots,N$.
According to this definition, the Hamiltonian considered in this paper is assumed to belong to the class of $k$-local and $g$-extensive operators.

Let $A$ be a $k_A$-local and $g_A$-extensive operator and let $B$ be a $k_B$-local and $g_B$-extensive operator.  
Then we readily find that the commutator of $A$ and $B$ is $(k_A+k_B)$-local and $[2(k_A+k_B)g_Ag_B]$-extensive.
Using this iteratively leads to that the multiple commutator in Eq.~(\ref{eq:FM}) is $(n+1)k$-local and $\left[g(2gk)^n(n+1)!\right]$-extensive.
From this we find that $\Omega_n$ is $(n+1)k$-local and $g_n$-extensive with 
\begin{eqnarray}
g_n &=&\frac{(2gk)^nn!}{n+1}g \,  \label{eq:gn}.
\end{eqnarray}

We can derive useful inequality for the (multiple) commutators of $k_A$-local and $g_A$-extensive operator $A$ and $k_B$-local operator $B=\sum_{|X|\leq k_B}b_X$:
\beq
\|[A,B]\|  \leq  2g_Ak_B\overline{B} \, ,
\label{eq:inequality}
\eeq
where $\overline{B}\equiv\sum_{|X|\leq k_B}\|b_X\|$.
More generally, we can show that for any $k_{A_i}$-local and $g_{A_i}$-extensive operators $\{A_i\}$,
\beq
\|[A_n,[A_{n-1},\dots,[A_1,B]\dots]]\|\leq \overline{B}\, \prod_{i=1}^n2g_{A_i}K_i \, 
,
\label{eq:multi_inequality}
\eeq
where $K_i\equiv k_B+\sum_{j=1}^{i-1}k_{A_i}$. 

By applying this inequality to Eq.~(\ref{eq:FM}), it is shown that the coefficient of the FM expansion $\Omega_n$ can be decomposed as $\Omega_n=\sum_{X:|X|\leq (n+1)k}w_X$, where $w_X$ is an operator acting on the sites in $X$, with
\beq
\overline{\Omega_n}T^n\equiv\sum_{X:|X|\leq (n+1)k}\|w_X\|T^n\leq 2gN_V\frac{(2gkT)^nn!}{(n+1)^2}.
\label{eq:Omega}
\eeq
Obviously $\|\Omega_n\|\leq\overline{\Omega_n}$ and thus Eq.~(\ref{eq:bound_Omega}) is also derived.

\bigskip
\noindent
\subsection{Time evolution of local operators over single period}

In order to prove the theorem, we first consider the time evolution of a local operator $O$ in the single period.
The exact time evolution of the operator $O$ in the Heisenberg picture is given by
\beq
O(t)=U^{\dagger}(t)OU(t)=\overline{\mathcal{T}}e^{i\int_0^tdt'L(t')}O,
\label{eq:O(t)}
\eeq
where $L(t)(\cdot)=[H(t),\cdot]$ is the Liouville operator.
We define the approximate time evolution under the truncated Floquet Hamiltonian $H_F^{(n_0)}$ as
\beq
\tilde{O}^{(n_0)}(t)=e^{iH_F^{(n_0)}t}Oe^{-iH_F^{(n_0)}t}=e^{iL_F^{(n_0)}t}O,
\label{eq:O(t)'}
\eeq
where $L_F^{(n_0)}(\cdot)=[H_F^{(n_0)},\cdot]$.
We can show the following lemma:

\bigskip
\noindent
\textbf{Lemma}.
\textit{Assume that $H_0$, $V(t)$, and $H(t)$ are $k$-local and $g$-extensive.
Then, for an arbitrary $(I+1)k$-local operator $O=\sum_{|X|\leq (I+1)k}o_X$ and the period $T<1/(8gk)$, the following inequality holds:
\beq
\|O(T)-\tilde{O}^{(n_0)}(T)\|\leq 16gk\overline{O}2^{-(n_0-I)}T,
\label{eq:theorem}
\eeq
where $\overline{O}\equiv\sum_{|X|\leq (I+1)k}\|o_X\|$ and $n_0=\lfloor 1/(8gkT)-1\rfloor$.
In particular, for $O=H_0$, the following stronger bound exists:
\beq
\|H_0(T)-\tilde{H}_0^{(n_0)}(T)\|\leq 8g^2k2^{-n_0}N_VT,
\label{eq:theorem_H0}
\eeq
where $N_V$ is the number of sites subjected to the periodic driving.}
\bigskip

Note that Eqs.~(\ref{eq:O(t)}) and (\ref{eq:O(t)'}) are rewritten as follows
\begin{eqnarray} 
\begin{array}{l}
~~~~\, O(T)=\overline{\mathcal{T}}e^{i\int_0^TdtL(t)}O = \sum_{n=0}^{\infty}T^n\mathcal{A}_n {\cal O}\, , \\ [5pt] 
\tilde{O}^{(n_0)}(T) = e^{iL_F^{(n_0)}T}O =\sum_{n=0}^{\infty}T^n\mathcal{\tilde{A}}_n^{(n_0)}  {\cal O} \, , 
\end{array} \label{heise2}
\end{eqnarray}
where $L(t) =[H(t),\cdot ] $ and $L_F^{(n_0)} = [H_F^{(n_0)},\cdot ]$, and $\mathcal{A}_n$ is the coefficient in the Dyson expansion given as
\begin{align}
\mathcal{A}_n=\frac{i^n}{T^n}\int_0^Tdt_n\dots\int_{t_2}^Tdt_1
\,  L(t_n)\dots L(t_1).
\label{eq:Dyson}
\end{align}
The coefficient $\mathcal{\tilde{A}}_n^{(n_0)}$ is given by
\begin{eqnarray}
\!\!\mathcal{\tilde{A}}_n^{(n_0)}&=&\sum_{r=0}^n\sum_{\substack{\{l_i\}_{i=1}^r \\ 0\leq l_i\leq n_0}}
\!\!\frac{i^r}{r!} \chi\left(\sum_{i=1}^r(l_i+1)=n\right)
L_{l_1}\dots L_{l_r}, ~~~~~
\label{eq:Taylor}
\end{eqnarray}
where $L_l\equiv[\Omega_l,\cdot]$ and $\chi$ is the indicator function defined by $\chi({\rm True})=1$ and $\chi({\rm False})=0$. The truncated FM expansion is exact up to $\mathcal{O}(T^{n_0})$ in the sense that
\beq
\mathcal{A}_n=\mathcal{\tilde{A}}_n^{(n_0)} \quad \text{for all $n\leq n_0$}.
\eeq
By using this fact, we have
\begin{align}
\|O(T)-\tilde{O}(T)\| &\leq \sum_{n=n_0+1}^{\infty}T^n\left(\|\mathcal{A}_nO\|+\|\mathcal{\tilde{A}}_n^{(n_0)}O\|\right).
~~~~~~
\label{eq:O_bound}
\end{align}

We bound $\|\mathcal{A}_nO\|$ and $\|\tilde{\mathcal{A}}_n^{(n_0)}O\|$ from above, using Eq.~(\ref{eq:multi_inequality}).
First, we show from Eqs.~(\ref{eq:Dyson}) and (\ref{eq:multi_inequality}) that $\|\mathcal{A}_nO\|$ is bounded as
\begin{align}
\|\mathcal{A}_nO\| &\leq \frac{1}{n!}\sup_{0\leq t_1,\dots,t_n\leq T}\|L(t_n)\dots L(t_1)O\|
\leq (2gk)^n\frac{(n+I)!}{n!I!}\overline{O}.
\end{align}
By using $(n+I)!/[n!I!]\leq 2^{n+I}$ and $4gkT\leq 1/2$, we have
\beq
\sum_{n=n_0+1}^{\infty}T^n\|\mathcal{A}_nO\|\leq 8gkT2^{I-n_0}\overline{O}.
\label{eq:An}
\eeq
For $O=H_0$, we can give a stronger bound.
Because $L(t_1)H_0=[H(t_1),H_0]=[V(t_1),H_0]=-L_0V(t_1)$, where $L_0\equiv[H_0,\cdot]$, we have
\begin{eqnarray}
\|\mathcal{A}_nH_0\| &\leq & \frac{1}{n!}\sup_{0\leq t_1\dots,t_n\leq T}\|L(t_n)\dots L(t_2)L_0V(t_1)\|\, . \, ~~~~~
\end{eqnarray}
Because $V(t)$ is $k$-local, it is written as $V(t)=\sum_{|X|\leq k}v_X(t)$, where $v_X(t)$ is an operator acting non-trivially only on the domain $X$.
By almost the same calculation, we have
\beq
\sum_{n=n_0+1}^{\infty}T^n\|\mathcal{A}_nH_0\|\leq 4gkT2^{-n_0}V_0,
\label{eq:AnH0}
\eeq
where $V_0\equiv\sup_{0\leq t\leq T}\sum_{|X|\leq k}\|v_X(t)\|$.
Because $V(t)$ is $g$-extensive, $V_0\leq gN_V$ if the driving is applied to $N_V$ sites.

Next, we evaluate an upper bound of $\|\mathcal{\tilde{A}}_n^{(n_0)}O\|$,
\begin{align}
\|\mathcal{\tilde{A}}_n^{(n_0)}O\|\leq\sum_{r=0}^n\sum_{\substack{\{l_i\}_{i=1}^r \\ 0\leq l_i\leq n_0}}\frac{1}{r!}\chi\left(\sum_{i=1}^r(l_i+1)=n\right)\|L_{l_1}\dots L_{l_r}O\|.
\end{align}

By using Eqs.~(\ref{eq:multi_inequality}) and (\ref{eq:gn}) with $g_l\leq (2gkn_0)^lg$, because $l!\leq n_0^l$ for $l\leq n_0$, we obtain
\beq
\|L_{l_1}\dots L_{l_r}O\|\leq (2gk)^r(2gkn_0)^{n-r}\frac{(n+I)!}{(n+I-r)!}\overline{O}.
\eeq
By using $(n+I)!/[r!(n+I-r)!]\leq 2^{n+I}$ and $\sum_{\{l_i\}_{i=1}^r}\chi(\sum_{i=1}^r(l_i+1)=n)=(n-1)!/[(n-r)!(r-1)!]$, we obtain
\begin{align}
\|\mathcal{\tilde{A}}_n^{(n_0)}O\|&\leq 2^{n+I}\sum_{r=1}^n\frac{(n-1)!}{(n-r)!(r-1)!}(2gk)^r(2gkn_0)^{n-r}\overline{O}
=2^I[4gk(n_0+1)]^{n-1}4gk\overline{O}.
\end{align}
Because $n_0$ is the maximum integer not exceeding $1/(8gkT)-1$, we have $4gkT(n_0+1)\leq 1/2$ and thus
\beq
\sum_{n=n_0+1}^{\infty}T^n\|\mathcal{\tilde{A}}_n^{(n_0)}O\|\leq 8gkT2^{I-n_0}\overline{O}.
\label{eq:An2}
\eeq
By substituting Eqs.~(\ref{eq:An}) and (\ref{eq:An2}) into Eq.~(\ref{eq:O_bound}), we complete the proof of Lemma~(\ref{eq:theorem}).

For $O=H_0$, we have $L_{l_r}H_0=-L_0\Omega_{l_r}$ and thus $\|L_{l_1}\dots L_{l_r}H_0\|=\|L_{l_1}\dots L_{l_{r-1}}L_0\Omega_{l_r}\|$.
Clearly, $l_r=0$ has no contribution, and for $l_r\geq 1$, we can use Eq.~(\ref{eq:Omega}).
As $1\leq l_r\leq n_0$, we have $\overline{\Omega_{l_r}}\leq (2gkn_0)^{l_r}gN_V/2$.
Applying Eq.~(\ref{eq:multi_inequality}) and the upper bound of $\overline{\Omega_{l_r}}$,
we obtain
\beq
\sum_{n=n_0+1}^{\infty}T^n\|\tilde{\mathcal{A}}_n^{(n_0)}H_0\|\leq 4g^2kTN_V2^{-n_0}.
\label{eq:AnH02}
\eeq
Using Eqs.~(\ref{eq:AnH0}) and (\ref{eq:AnH02}), we obtain Lemma~(\ref{eq:theorem_H0}).

\bigskip
\noindent
\subsection{Proof of Theorem}

 From the relation $\tilde{H}_F^{(n_0)}(t)=H_F^{(n_0)}$, we note the inequality for any positive integer $m$:
\begin{eqnarray}
\|H_F^{(n_0)}(mT)-H_F^{(n_0)}\|&\leq & m\|H_F^{(n_0)}(T)-H_F^{(n_0)}\| \nonumber \\
&\leq & m\sum_{n=0}^{n_0}T^n\|\Omega_n(T)-\tilde{\Omega}_n^{(n_0)}(T)\| =m\|H_0(T)-\tilde{H}_0(T)\|+m\sum_{n=1}^{n_0}T^n\|\Omega_n(T)-\tilde{\Omega}_n^{(n_0)}(T)\|. ~~~~~~~~
\label{eq:diff_HF}
\end{eqnarray}
Applying Lemma and using Eq.~(\ref{eq:Omega}), we obtain
\begin{align}
&\|H_0(T)-\tilde{H}_0(T)\|\leq 8g^2k2^{-n_0}N_VT, 
\label{eq:bound_H0} \\
&\|\Omega_n(T)-\tilde{\Omega}_n^{(n_0)}(T)\|T^n\leq 16gk\overline{\Omega_n}2^{-(n_0-n)}T^{n+1}
\leq 32g^2k2^{-n_0}N_VT\frac{(4gkT)^nn!}{(n+1)^2}.
\end{align}
By using $n!\leq n_0^n\leq 1/(8gkT)^n$ and $1/(n+1)^2\leq 1/4$ for $1\leq n\leq n_0$, we have
\begin{align}
\sum_{n=1}^{n_0}\|\Omega_n(T)-\tilde{\Omega}_n^{(n_0)}(T)\|T^n &\leq 8g^2k2^{-n_0}N_VT\sum_{n=1}^{\infty}\left(\frac{1}{2}\right)^n
\nonumber \\
&=8g^2k2^{-n_0}N_VT.
\end{align}
By combining this and Eq.~(\ref{eq:bound_H0}) with Eq.~(\ref{eq:diff_HF}), we obtain the theorem
\beq
\|H_F^{(n_0)}(t)-H_F^{(n_0)}\|\leq 16g^2k2^{-n_0}N_Vt,
\label{eq:exp_slow}
\eeq
where $t=mT$.
This completes the proof of Theorem.


\begin{thebibliography}{49}%
\makeatletter
\providecommand \@ifxundefined [1]{%
 \@ifx{#1\undefined}
}%
\providecommand \@ifnum [1]{%
 \ifnum #1\expandafter \@firstoftwo
 \else \expandafter \@secondoftwo
 \fi
}%
\providecommand \@ifx [1]{%
 \ifx #1\expandafter \@firstoftwo
 \else \expandafter \@secondoftwo
 \fi
}%
\providecommand \natexlab [1]{#1}%
\providecommand \enquote  [1]{``#1''}%
\providecommand \bibnamefont  [1]{#1}%
\providecommand \bibfnamefont [1]{#1}%
\providecommand \citenamefont [1]{#1}%
\providecommand \href@noop [0]{\@secondoftwo}%
\providecommand \href [0]{\begingroup \@sanitize@url \@href}%
\providecommand \@href[1]{\@@startlink{#1}\@@href}%
\providecommand \@@href[1]{\endgroup#1\@@endlink}%
\providecommand \@sanitize@url [0]{\catcode `\\12\catcode `\$12\catcode
  `\&12\catcode `\#12\catcode `\^12\catcode `\_12\catcode `\%12\relax}%
\providecommand \@@startlink[1]{}%
\providecommand \@@endlink[0]{}%
\providecommand \url  [0]{\begingroup\@sanitize@url \@url }%
\providecommand \@url [1]{\endgroup\@href {#1}{\urlprefix }}%
\providecommand \urlprefix  [0]{URL }%
\providecommand \Eprint [0]{\href }%
\providecommand \doibase [0]{http://dx.doi.org/}%
\providecommand \selectlanguage [0]{\@gobble}%
\providecommand \bibinfo  [0]{\@secondoftwo}%
\providecommand \bibfield  [0]{\@secondoftwo}%
\providecommand \translation [1]{[#1]}%
\providecommand \BibitemOpen [0]{}%
\providecommand \bibitemStop [0]{}%
\providecommand \bibitemNoStop [0]{.\EOS\space}%
\providecommand \EOS [0]{\spacefactor3000\relax}%
\providecommand \BibitemShut  [1]{\csname bibitem#1\endcsname}%
\let\auto@bib@innerbib\@empty
\bibitem [{\citenamefont {Matsunaga}\ \emph {et~al.}(2014)\citenamefont
  {Matsunaga}, \citenamefont {Tsuji}, \citenamefont {Fujita}, \citenamefont
  {Sugioka}, \citenamefont {Makise}, \citenamefont {Uzawa}, \citenamefont
  {Terai}, \citenamefont {Wang}, \citenamefont {Aoki},\ and\ \citenamefont
  {Shimano}}]{matsunaga2014light}%
  \BibitemOpen
  \bibfield  {author} {\bibinfo {author} {\bibfnamefont {R.}~\bibnamefont
  {Matsunaga}}, \bibinfo {author} {\bibfnamefont {N.}~\bibnamefont {Tsuji}},
  \bibinfo {author} {\bibfnamefont {H.}~\bibnamefont {Fujita}}, \bibinfo
  {author} {\bibfnamefont {A.}~\bibnamefont {Sugioka}}, \bibinfo {author}
  {\bibfnamefont {K.}~\bibnamefont {Makise}}, \bibinfo {author} {\bibfnamefont
  {Y.}~\bibnamefont {Uzawa}}, \bibinfo {author} {\bibfnamefont
  {H.}~\bibnamefont {Terai}}, \bibinfo {author} {\bibfnamefont
  {Z.}~\bibnamefont {Wang}}, \bibinfo {author} {\bibfnamefont {H.}~\bibnamefont
  {Aoki}}, \ and\ \bibinfo {author} {\bibfnamefont {R.}~\bibnamefont
  {Shimano}},\ }\href {\doibase 10.1126/science.1254697} {\bibfield  {journal}
  {\bibinfo  {journal} {Science}\ }\textbf {\bibinfo {volume} {345}},\ \bibinfo
  {pages} {1145} (\bibinfo {year} {2014})}\BibitemShut {NoStop}%
\bibitem [{\citenamefont {Aidelsburger}\ \emph {et~al.}(2013)\citenamefont
  {Aidelsburger}, \citenamefont {Atala}, \citenamefont {Lohse}, \citenamefont
  {Barreiro}, \citenamefont {Paredes},\ and\ \citenamefont
  {Bloch}}]{Aidelsburger2013}%
  \BibitemOpen
  \bibfield  {author} {\bibinfo {author} {\bibfnamefont {M.}~\bibnamefont
  {Aidelsburger}}, \bibinfo {author} {\bibfnamefont {M.}~\bibnamefont {Atala}},
  \bibinfo {author} {\bibfnamefont {M.}~\bibnamefont {Lohse}}, \bibinfo
  {author} {\bibfnamefont {J.~T.}\ \bibnamefont {Barreiro}}, \bibinfo {author}
  {\bibfnamefont {B.}~\bibnamefont {Paredes}}, \ and\ \bibinfo {author}
  {\bibfnamefont {I.}~\bibnamefont {Bloch}},\ }\href {\doibase
  10.1103/PhysRevLett.111.185301} {\bibfield  {journal} {\bibinfo  {journal}
  {Phys. Rev. Lett.}\ }\textbf {\bibinfo {volume} {111}},\ \bibinfo {pages}
  {185301} (\bibinfo {year} {2013})}\BibitemShut {NoStop}%
\bibitem [{\citenamefont {Atala}\ \emph {et~al.}(2013)\citenamefont {Atala},
  \citenamefont {Aidelsburger}, \citenamefont {Barreiro}, \citenamefont
  {Abanin}, \citenamefont {Kitagawa}, \citenamefont {Demler},\ and\
  \citenamefont {Bloch}}]{Atala2013}%
  \BibitemOpen
  \bibfield  {author} {\bibinfo {author} {\bibfnamefont {M.}~\bibnamefont
  {Atala}}, \bibinfo {author} {\bibfnamefont {M.}~\bibnamefont {Aidelsburger}},
  \bibinfo {author} {\bibfnamefont {J.~T.}\ \bibnamefont {Barreiro}}, \bibinfo
  {author} {\bibfnamefont {D.}~\bibnamefont {Abanin}}, \bibinfo {author}
  {\bibfnamefont {T.}~\bibnamefont {Kitagawa}}, \bibinfo {author}
  {\bibfnamefont {E.}~\bibnamefont {Demler}}, \ and\ \bibinfo {author}
  {\bibfnamefont {I.}~\bibnamefont {Bloch}},\ }\href {\doibase
  10.1038/nphys2790} {\bibfield  {journal} {\bibinfo  {journal} {Nature
  Physics}\ }\textbf {\bibinfo {volume} {9}},\ \bibinfo {pages} {795} (\bibinfo
  {year} {2013})}\BibitemShut {NoStop}%
\bibitem [{\citenamefont {Jotzu}\ \emph {et~al.}(2014)\citenamefont {Jotzu},
  \citenamefont {Messer}, \citenamefont {Desbuquois}, \citenamefont {Lebrat},
  \citenamefont {Uehlinger}, \citenamefont {Greif},\ and\ \citenamefont
  {Esslinger}}]{jotzu2014experimental}%
  \BibitemOpen
  \bibfield  {author} {\bibinfo {author} {\bibfnamefont {G.}~\bibnamefont
  {Jotzu}}, \bibinfo {author} {\bibfnamefont {M.}~\bibnamefont {Messer}},
  \bibinfo {author} {\bibfnamefont {R.}~\bibnamefont {Desbuquois}}, \bibinfo
  {author} {\bibfnamefont {M.}~\bibnamefont {Lebrat}}, \bibinfo {author}
  {\bibfnamefont {T.}~\bibnamefont {Uehlinger}}, \bibinfo {author}
  {\bibfnamefont {D.}~\bibnamefont {Greif}}, \ and\ \bibinfo {author}
  {\bibfnamefont {T.}~\bibnamefont {Esslinger}},\ }\href {\doibase
  10.1038/nature13915} {\bibfield  {journal} {\bibinfo  {journal} {Nature}\
  }\textbf {\bibinfo {volume} {515}},\ \bibinfo {pages} {237} (\bibinfo {year}
  {2014})}\BibitemShut {NoStop}%
\bibitem [{\citenamefont {Aidelsburger}\ \emph {et~al.}(2015)\citenamefont
  {Aidelsburger}, \citenamefont {Lohse}, \citenamefont {Schweizer},
  \citenamefont {Atala}, \citenamefont {Barreiro}, \citenamefont {Nascimbene},
  \citenamefont {Cooper}, \citenamefont {Bloch},\ and\ \citenamefont
  {Goldman}}]{Aidelsburger2015}%
  \BibitemOpen
  \bibfield  {author} {\bibinfo {author} {\bibfnamefont {M.}~\bibnamefont
  {Aidelsburger}}, \bibinfo {author} {\bibfnamefont {M.}~\bibnamefont {Lohse}},
  \bibinfo {author} {\bibfnamefont {C.}~\bibnamefont {Schweizer}}, \bibinfo
  {author} {\bibfnamefont {M.}~\bibnamefont {Atala}}, \bibinfo {author}
  {\bibfnamefont {J.~T.}\ \bibnamefont {Barreiro}}, \bibinfo {author}
  {\bibfnamefont {S.}~\bibnamefont {Nascimbene}}, \bibinfo {author}
  {\bibfnamefont {N.}~\bibnamefont {Cooper}}, \bibinfo {author} {\bibfnamefont
  {I.}~\bibnamefont {Bloch}}, \ and\ \bibinfo {author} {\bibfnamefont
  {N.}~\bibnamefont {Goldman}},\ }\href {\doibase 10.1038/nphys3171} {\bibfield
   {journal} {\bibinfo  {journal} {Nature Physics}\ }\textbf {\bibinfo {volume}
  {11}},\ \bibinfo {pages} {162} (\bibinfo {year} {2015})}\BibitemShut
  {NoStop}%
\bibitem [{\citenamefont {Dunlap}\ and\ \citenamefont
  {Kenkre}(1986)}]{Dunlap-Kenkre1986}%
  \BibitemOpen
  \bibfield  {author} {\bibinfo {author} {\bibfnamefont {D.~H.}\ \bibnamefont
  {Dunlap}}\ and\ \bibinfo {author} {\bibfnamefont {V.~M.}\ \bibnamefont
  {Kenkre}},\ }\href {\doibase 10.1103/PhysRevB.34.3625} {\bibfield  {journal}
  {\bibinfo  {journal} {Phys. Rev. B}\ }\textbf {\bibinfo {volume} {34}},\
  \bibinfo {pages} {3625} (\bibinfo {year} {1986})}\BibitemShut {NoStop}%
\bibitem [{\citenamefont {Grifoni}\ and\ \citenamefont
  {H{\"a}nggi}(1998)}]{Grifoni_review1998}%
  \BibitemOpen
  \bibfield  {author} {\bibinfo {author} {\bibfnamefont {M.}~\bibnamefont
  {Grifoni}}\ and\ \bibinfo {author} {\bibfnamefont {P.}~\bibnamefont
  {H{\"a}nggi}},\ }\href {\doibase 10.1016/S0370-1573(98)00022-2} {\bibfield
  {journal} {\bibinfo  {journal} {Phys. Rep.}\ }\textbf {\bibinfo {volume}
  {304}},\ \bibinfo {pages} {229} (\bibinfo {year} {1998})}\BibitemShut
  {NoStop}%
\bibitem [{\citenamefont {Kayanuma}\ and\ \citenamefont
  {Saito}(2008)}]{Kayanuma-Saito2008}%
  \BibitemOpen
  \bibfield  {author} {\bibinfo {author} {\bibfnamefont {Y.}~\bibnamefont
  {Kayanuma}}\ and\ \bibinfo {author} {\bibfnamefont {K.}~\bibnamefont
  {Saito}},\ }\href {\doibase 10.1103/PhysRevA.77.010101} {\bibfield  {journal}
  {\bibinfo  {journal} {Phys. Rev. A}\ }\textbf {\bibinfo {volume} {77}},\
  \bibinfo {pages} {010101} (\bibinfo {year} {2008})}\BibitemShut {NoStop}%
\bibitem [{\citenamefont {Grossmann}\ \emph {et~al.}(1991)\citenamefont
  {Grossmann}, \citenamefont {Dittrich}, \citenamefont {Jung},\ and\
  \citenamefont {H\"anggi}}]{Grossmann1991}%
  \BibitemOpen
  \bibfield  {author} {\bibinfo {author} {\bibfnamefont {F.}~\bibnamefont
  {Grossmann}}, \bibinfo {author} {\bibfnamefont {T.}~\bibnamefont {Dittrich}},
  \bibinfo {author} {\bibfnamefont {P.}~\bibnamefont {Jung}}, \ and\ \bibinfo
  {author} {\bibfnamefont {P.}~\bibnamefont {H\"anggi}},\ }\href {\doibase
  10.1103/PhysRevLett.67.516} {\bibfield  {journal} {\bibinfo  {journal} {Phys.
  Rev. Lett.}\ }\textbf {\bibinfo {volume} {67}},\ \bibinfo {pages} {516}
  (\bibinfo {year} {1991})}\BibitemShut {NoStop}%
\bibitem [{\citenamefont {Das}(2010)}]{Das2010}%
  \BibitemOpen
  \bibfield  {author} {\bibinfo {author} {\bibfnamefont {A.}~\bibnamefont
  {Das}},\ }\href {\doibase 10.1103/PhysRevB.82.172402} {\bibfield  {journal}
  {\bibinfo  {journal} {Phys. Rev. B}\ }\textbf {\bibinfo {volume} {82}},\
  \bibinfo {pages} {172402} (\bibinfo {year} {2010})}\BibitemShut {NoStop}%
\bibitem [{\citenamefont {Hegde}\ \emph {et~al.}(2014)\citenamefont {Hegde},
  \citenamefont {Katiyar}, \citenamefont {Mahesh},\ and\ \citenamefont
  {Das}}]{Hegde2014}%
  \BibitemOpen
  \bibfield  {author} {\bibinfo {author} {\bibfnamefont {S.~S.}\ \bibnamefont
  {Hegde}}, \bibinfo {author} {\bibfnamefont {H.}~\bibnamefont {Katiyar}},
  \bibinfo {author} {\bibfnamefont {T.~S.}\ \bibnamefont {Mahesh}}, \ and\
  \bibinfo {author} {\bibfnamefont {A.}~\bibnamefont {Das}},\ }\href {\doibase
  10.1103/PhysRevB.90.174407} {\bibfield  {journal} {\bibinfo  {journal} {Phys.
  Rev. B}\ }\textbf {\bibinfo {volume} {90}},\ \bibinfo {pages} {174407}
  (\bibinfo {year} {2014})}\BibitemShut {NoStop}%
\bibitem [{\citenamefont {Prosen}\ and\ \citenamefont
  {Ilievski}(2011)}]{Prosen-Ilievski2011}%
  \BibitemOpen
  \bibfield  {author} {\bibinfo {author} {\bibfnamefont {T.}~\bibnamefont
  {Prosen}}\ and\ \bibinfo {author} {\bibfnamefont {E.}~\bibnamefont
  {Ilievski}},\ }\href {\doibase 10.1103/PhysRevLett.107.060403} {\bibfield
  {journal} {\bibinfo  {journal} {Phys. Rev. Lett.}\ }\textbf {\bibinfo
  {volume} {107}},\ \bibinfo {pages} {060403} (\bibinfo {year}
  {2011})}\BibitemShut {NoStop}%
\bibitem [{\citenamefont {Bastidas}\ \emph {et~al.}(2012)\citenamefont
  {Bastidas}, \citenamefont {Emary}, \citenamefont {Regler},\ and\
  \citenamefont {Brandes}}]{Bastidas2012}%
  \BibitemOpen
  \bibfield  {author} {\bibinfo {author} {\bibfnamefont {V.~M.}\ \bibnamefont
  {Bastidas}}, \bibinfo {author} {\bibfnamefont {C.}~\bibnamefont {Emary}},
  \bibinfo {author} {\bibfnamefont {B.}~\bibnamefont {Regler}}, \ and\ \bibinfo
  {author} {\bibfnamefont {T.}~\bibnamefont {Brandes}},\ }\href {\doibase
  10.1103/PhysRevLett.108.043003} {\bibfield  {journal} {\bibinfo  {journal}
  {Phys. Rev. Lett.}\ }\textbf {\bibinfo {volume} {108}},\ \bibinfo {pages}
  {043003} (\bibinfo {year} {2012})}\BibitemShut {NoStop}%
\bibitem [{\citenamefont {Shirai}\ \emph {et~al.}(2014)\citenamefont {Shirai},
  \citenamefont {Mori},\ and\ \citenamefont {Miyashita}}]{Shirai2014a}%
  \BibitemOpen
  \bibfield  {author} {\bibinfo {author} {\bibfnamefont {T.}~\bibnamefont
  {Shirai}}, \bibinfo {author} {\bibfnamefont {T.}~\bibnamefont {Mori}}, \ and\
  \bibinfo {author} {\bibfnamefont {S.}~\bibnamefont {Miyashita}},\ }\href
  {\doibase 10.1088/0953-4075/47/2/025501} {\bibfield  {journal} {\bibinfo
  {journal} {J. Phys. B}\ }\textbf {\bibinfo {volume} {47}},\ \bibinfo {pages}
  {025501} (\bibinfo {year} {2014})}\BibitemShut {NoStop}%
\bibitem [{\citenamefont {Russomanno}\ \emph {et~al.}(2012)\citenamefont
  {Russomanno}, \citenamefont {Silva},\ and\ \citenamefont
  {Santoro}}]{Russomanno2012}%
  \BibitemOpen
  \bibfield  {author} {\bibinfo {author} {\bibfnamefont {A.}~\bibnamefont
  {Russomanno}}, \bibinfo {author} {\bibfnamefont {A.}~\bibnamefont {Silva}}, \
  and\ \bibinfo {author} {\bibfnamefont {G.~E.}\ \bibnamefont {Santoro}},\
  }\href {\doibase 10.1103/PhysRevLett.109.257201} {\bibfield  {journal}
  {\bibinfo  {journal} {Phys. Rev. Lett.}\ }\textbf {\bibinfo {volume} {109}},\
  \bibinfo {pages} {257201} (\bibinfo {year} {2012})}\BibitemShut {NoStop}%
\bibitem [{\citenamefont {D'Alessio}\ and\ \citenamefont
  {Polkovnikov}(2013)}]{D'Alessio2013}%
  \BibitemOpen
  \bibfield  {author} {\bibinfo {author} {\bibfnamefont {L.}~\bibnamefont
  {D'Alessio}}\ and\ \bibinfo {author} {\bibfnamefont {A.}~\bibnamefont
  {Polkovnikov}},\ }\href {\doibase 10.1016/j.aop.2013.02.011} {\bibfield
  {journal} {\bibinfo  {journal} {Ann. Phys.}\ }\textbf {\bibinfo {volume}
  {333}},\ \bibinfo {pages} {19} (\bibinfo {year} {2013})}\BibitemShut
  {NoStop}%
\bibitem [{\citenamefont {Lazarides}\ \emph
  {et~al.}(2014{\natexlab{a}})\citenamefont {Lazarides}, \citenamefont {Das},\
  and\ \citenamefont {Moessner}}]{Lazarides2014a}%
  \BibitemOpen
  \bibfield  {author} {\bibinfo {author} {\bibfnamefont {A.}~\bibnamefont
  {Lazarides}}, \bibinfo {author} {\bibfnamefont {A.}~\bibnamefont {Das}}, \
  and\ \bibinfo {author} {\bibfnamefont {R.}~\bibnamefont {Moessner}},\ }\href
  {\doibase 10.1103/PhysRevLett.112.150401} {\bibfield  {journal} {\bibinfo
  {journal} {Phys. Rev. Lett.}\ }\textbf {\bibinfo {volume} {112}},\ \bibinfo
  {pages} {150401} (\bibinfo {year} {2014}{\natexlab{a}})}\BibitemShut
  {NoStop}%
\bibitem [{\citenamefont {Lazarides}\ \emph {et~al.}(2015)\citenamefont
  {Lazarides}, \citenamefont {Das},\ and\ \citenamefont
  {Moessner}}]{Lazarides2015}%
  \BibitemOpen
  \bibfield  {author} {\bibinfo {author} {\bibfnamefont {A.}~\bibnamefont
  {Lazarides}}, \bibinfo {author} {\bibfnamefont {A.}~\bibnamefont {Das}}, \
  and\ \bibinfo {author} {\bibfnamefont {R.}~\bibnamefont {Moessner}},\ }\href
  {\doibase 10.1103/PhysRevLett.115.030402} {\bibfield  {journal} {\bibinfo
  {journal} {Phys. Rev. Lett.}\ }\textbf {\bibinfo {volume} {115}},\ \bibinfo
  {pages} {030402} (\bibinfo {year} {2015})}\BibitemShut {NoStop}%
\bibitem [{\citenamefont {Ponte}\ \emph {et~al.}(2015)\citenamefont {Ponte},
  \citenamefont {Chandran}, \citenamefont {Papi{\'c}},\ and\ \citenamefont
  {Abanin}}]{Ponte2015}%
  \BibitemOpen
  \bibfield  {author} {\bibinfo {author} {\bibfnamefont {P.}~\bibnamefont
  {Ponte}}, \bibinfo {author} {\bibfnamefont {A.}~\bibnamefont {Chandran}},
  \bibinfo {author} {\bibfnamefont {Z.}~\bibnamefont {Papi{\'c}}}, \ and\
  \bibinfo {author} {\bibfnamefont {D.~A.}\ \bibnamefont {Abanin}},\ }\href
  {\doibase 10.1016/j.aop.2014.11.008} {\bibfield  {journal} {\bibinfo
  {journal} {Ann. Phys.}\ }\textbf {\bibinfo {volume} {353}},\ \bibinfo {pages}
  {196} (\bibinfo {year} {2015})}\BibitemShut {NoStop}%
\bibitem [{\citenamefont {Deutsch}(1991)}]{Deutsch1991}%
  \BibitemOpen
  \bibfield  {author} {\bibinfo {author} {\bibfnamefont {J.~M.}\ \bibnamefont
  {Deutsch}},\ }\href {\doibase 10.1103/PhysRevA.43.2046} {\bibfield  {journal}
  {\bibinfo  {journal} {Phys. Rev. A}\ }\textbf {\bibinfo {volume} {43}},\
  \bibinfo {pages} {2046} (\bibinfo {year} {1991})}\BibitemShut {NoStop}%
\bibitem [{\citenamefont {Srednicki}(1994)}]{Srednicki1994}%
  \BibitemOpen
  \bibfield  {author} {\bibinfo {author} {\bibfnamefont {M.}~\bibnamefont
  {Srednicki}},\ }\href {\doibase 10.1103/PhysRevE.50.888} {\bibfield
  {journal} {\bibinfo  {journal} {Phys. Rev. E}\ }\textbf {\bibinfo {volume}
  {50}},\ \bibinfo {pages} {888} (\bibinfo {year} {1994})}\BibitemShut
  {NoStop}%
\bibitem [{\citenamefont {Tasaki}(1998)}]{Tasaki1998}%
  \BibitemOpen
  \bibfield  {author} {\bibinfo {author} {\bibfnamefont {H.}~\bibnamefont
  {Tasaki}},\ }\href {\doibase 10.1103/PhysRevLett.80.1373} {\bibfield
  {journal} {\bibinfo  {journal} {Phys. Rev. Lett.}\ }\textbf {\bibinfo
  {volume} {80}},\ \bibinfo {pages} {1373} (\bibinfo {year}
  {1998})}\BibitemShut {NoStop}%
\bibitem [{\citenamefont {Rigol}\ \emph {et~al.}(2008)\citenamefont {Rigol},
  \citenamefont {Dunjko},\ and\ \citenamefont {Olshanii}}]{Rigol2008}%
  \BibitemOpen
  \bibfield  {author} {\bibinfo {author} {\bibfnamefont {M.}~\bibnamefont
  {Rigol}}, \bibinfo {author} {\bibfnamefont {V.}~\bibnamefont {Dunjko}}, \
  and\ \bibinfo {author} {\bibfnamefont {M.}~\bibnamefont {Olshanii}},\ }\href
  {\doibase 10.1038/nature06838} {\bibfield  {journal} {\bibinfo  {journal}
  {Nature}\ }\textbf {\bibinfo {volume} {452}},\ \bibinfo {pages} {854}
  (\bibinfo {year} {2008})}\BibitemShut {NoStop}%
\bibitem [{\citenamefont {Polkovnikov}\ \emph {et~al.}(2011)\citenamefont
  {Polkovnikov}, \citenamefont {Sengupta}, \citenamefont {Silva},\ and\
  \citenamefont {Vengalattore}}]{Polkovnikov_review2011}%
  \BibitemOpen
  \bibfield  {author} {\bibinfo {author} {\bibfnamefont {A.}~\bibnamefont
  {Polkovnikov}}, \bibinfo {author} {\bibfnamefont {K.}~\bibnamefont
  {Sengupta}}, \bibinfo {author} {\bibfnamefont {A.}~\bibnamefont {Silva}}, \
  and\ \bibinfo {author} {\bibfnamefont {M.}~\bibnamefont {Vengalattore}},\
  }\href {\doibase 10.1103/RevModPhys.83.863} {\bibfield  {journal} {\bibinfo
  {journal} {Rev. Mod. Phys.}\ }\textbf {\bibinfo {volume} {83}},\ \bibinfo
  {pages} {863} (\bibinfo {year} {2011})}\BibitemShut {NoStop}%
\bibitem [{\citenamefont {Sato}\ \emph {et~al.}(2012)\citenamefont {Sato},
  \citenamefont {Kanamoto}, \citenamefont {Kaminishi},\ and\ \citenamefont
  {Deguchi}}]{Sato2012}%
  \BibitemOpen
  \bibfield  {author} {\bibinfo {author} {\bibfnamefont {J.}~\bibnamefont
  {Sato}}, \bibinfo {author} {\bibfnamefont {R.}~\bibnamefont {Kanamoto}},
  \bibinfo {author} {\bibfnamefont {E.}~\bibnamefont {Kaminishi}}, \ and\
  \bibinfo {author} {\bibfnamefont {T.}~\bibnamefont {Deguchi}},\ }\href
  {\doibase 10.1103/PhysRevLett.108.110401} {\bibfield  {journal} {\bibinfo
  {journal} {Phys. Rev. Lett.}\ }\textbf {\bibinfo {volume} {108}},\ \bibinfo
  {pages} {110401} (\bibinfo {year} {2012})}\BibitemShut {NoStop}%
\bibitem [{\citenamefont {Neumann}(1929)}]{Neumann1929}%
  \BibitemOpen
  \bibfield  {author} {\bibinfo {author} {\bibfnamefont {J.~v.}\ \bibnamefont
  {Neumann}},\ }\href {\doibase 10.1007/BF01339852} {\bibfield  {journal}
  {\bibinfo  {journal} {Z. Phys.}\ }\textbf {\bibinfo {volume} {57}},\ \bibinfo
  {pages} {30} (\bibinfo {year} {1929})}\BibitemShut {NoStop}%
\bibitem [{\citenamefont {Popescu}\ \emph {et~al.}(2006)\citenamefont
  {Popescu}, \citenamefont {Short},\ and\ \citenamefont
  {Winter}}]{Popescu2006}%
  \BibitemOpen
  \bibfield  {author} {\bibinfo {author} {\bibfnamefont {S.}~\bibnamefont
  {Popescu}}, \bibinfo {author} {\bibfnamefont {A.~J.}\ \bibnamefont {Short}},
  \ and\ \bibinfo {author} {\bibfnamefont {A.}~\bibnamefont {Winter}},\ }\href
  {\doibase 10.1038/nphys444} {\bibfield  {journal} {\bibinfo  {journal}
  {Nature Phys.}\ }\textbf {\bibinfo {volume} {2}},\ \bibinfo {pages} {754}
  (\bibinfo {year} {2006})}\BibitemShut {NoStop}%
\bibitem [{\citenamefont {Bloch}\ \emph {et~al.}(2008)\citenamefont {Bloch},
  \citenamefont {Dalibard},\ and\ \citenamefont {Zwerger}}]{Bloch_review2008}%
  \BibitemOpen
  \bibfield  {author} {\bibinfo {author} {\bibfnamefont {I.}~\bibnamefont
  {Bloch}}, \bibinfo {author} {\bibfnamefont {J.}~\bibnamefont {Dalibard}}, \
  and\ \bibinfo {author} {\bibfnamefont {W.}~\bibnamefont {Zwerger}},\ }\href
  {\doibase 10.1103/RevModPhys.80.885} {\bibfield  {journal} {\bibinfo
  {journal} {Rev. Mod. Phys.}\ }\textbf {\bibinfo {volume} {80}},\ \bibinfo
  {pages} {885} (\bibinfo {year} {2008})}\BibitemShut {NoStop}%
\bibitem [{\citenamefont {D'Alessio}\ and\ \citenamefont
  {Rigol}(2014)}]{D'Alessio2014}%
  \BibitemOpen
  \bibfield  {author} {\bibinfo {author} {\bibfnamefont {L.}~\bibnamefont
  {D'Alessio}}\ and\ \bibinfo {author} {\bibfnamefont {M.}~\bibnamefont
  {Rigol}},\ }\href {\doibase 10.1103/PhysRevX.4.041048} {\bibfield  {journal}
  {\bibinfo  {journal} {Phys. Rev. X}\ }\textbf {\bibinfo {volume} {4}},\
  \bibinfo {pages} {041048} (\bibinfo {year} {2014})}\BibitemShut {NoStop}%
\bibitem [{\citenamefont {Lazarides}\ \emph
  {et~al.}(2014{\natexlab{b}})\citenamefont {Lazarides}, \citenamefont {Das},\
  and\ \citenamefont {Moessner}}]{Lazarides2014b}%
  \BibitemOpen
  \bibfield  {author} {\bibinfo {author} {\bibfnamefont {A.}~\bibnamefont
  {Lazarides}}, \bibinfo {author} {\bibfnamefont {A.}~\bibnamefont {Das}}, \
  and\ \bibinfo {author} {\bibfnamefont {R.}~\bibnamefont {Moessner}},\ }\href
  {\doibase 10.1103/PhysRevE.90.012110} {\bibfield  {journal} {\bibinfo
  {journal} {Phys. Rev. E}\ }\textbf {\bibinfo {volume} {90}},\ \bibinfo
  {pages} {012110} (\bibinfo {year} {2014}{\natexlab{b}})}\BibitemShut
  {NoStop}%
\bibitem [{\citenamefont {Kim}\ \emph {et~al.}(2014)\citenamefont {Kim},
  \citenamefont {Ikeda},\ and\ \citenamefont {Huse}}]{Kim2014}%
  \BibitemOpen
  \bibfield  {author} {\bibinfo {author} {\bibfnamefont {H.}~\bibnamefont
  {Kim}}, \bibinfo {author} {\bibfnamefont {T.~N.}\ \bibnamefont {Ikeda}}, \
  and\ \bibinfo {author} {\bibfnamefont {D.~A.}\ \bibnamefont {Huse}},\ }\href
  {\doibase 10.1103/PhysRevE.90.052105} {\bibfield  {journal} {\bibinfo
  {journal} {Phys. Rev. E}\ }\textbf {\bibinfo {volume} {90}},\ \bibinfo
  {pages} {052105} (\bibinfo {year} {2014})}\BibitemShut {NoStop}%
\bibitem [{\citenamefont {Blanes}\ \emph {et~al.}(2009)\citenamefont {Blanes},
  \citenamefont {Casas}, \citenamefont {Oteo},\ and\ \citenamefont
  {Ros}}]{Blanes_review2009}%
  \BibitemOpen
  \bibfield  {author} {\bibinfo {author} {\bibfnamefont {S.}~\bibnamefont
  {Blanes}}, \bibinfo {author} {\bibfnamefont {F.}~\bibnamefont {Casas}},
  \bibinfo {author} {\bibfnamefont {J.}~\bibnamefont {Oteo}}, \ and\ \bibinfo
  {author} {\bibfnamefont {J.}~\bibnamefont {Ros}},\ }\href {\doibase
  10.1016/j.physrep.2008.11.001} {\bibfield  {journal} {\bibinfo  {journal}
  {Phys. Rep.}\ }\textbf {\bibinfo {volume} {470}},\ \bibinfo {pages} {151}
  (\bibinfo {year} {2009})}\BibitemShut {NoStop}%
\bibitem [{\citenamefont {Bukov}\ \emph
  {et~al.}(2015{\natexlab{a}})\citenamefont {Bukov}, \citenamefont
  {D'Alessio},\ and\ \citenamefont {Polkovnikov}}]{Bukov_review2015}%
  \BibitemOpen
  \bibfield  {author} {\bibinfo {author} {\bibfnamefont {M.}~\bibnamefont
  {Bukov}}, \bibinfo {author} {\bibfnamefont {L.}~\bibnamefont {D'Alessio}}, \
  and\ \bibinfo {author} {\bibfnamefont {A.}~\bibnamefont {Polkovnikov}},\
  }\href {\doibase 10.1080/00018732.2015.1055918} {\bibfield  {journal}
  {\bibinfo  {journal} {Advances in Physics}\ }\textbf {\bibinfo {volume}
  {64}},\ \bibinfo {pages} {139} (\bibinfo {year}
  {2015}{\natexlab{a}})}\BibitemShut {NoStop}%
\bibitem [{\citenamefont {Mori}(2015)}]{Mori2015_Floquet}%
  \BibitemOpen
  \bibfield  {author} {\bibinfo {author} {\bibfnamefont {T.}~\bibnamefont
  {Mori}},\ }\href {\doibase 10.1103/PhysRevA.91.020101} {\bibfield  {journal}
  {\bibinfo  {journal} {Phys. Rev. A}\ }\textbf {\bibinfo {volume} {91}},\
  \bibinfo {pages} {020101} (\bibinfo {year} {2015})}\BibitemShut {NoStop}%
\bibitem [{\citenamefont {Kuwahara}\ \emph {et~al.}(2016)\citenamefont
  {Kuwahara}, \citenamefont {Mori},\ and\ \citenamefont
  {Saito}}]{Kuwahara2016Floquet}%
  \BibitemOpen
  \bibfield  {author} {\bibinfo {author} {\bibfnamefont {T.}~\bibnamefont
  {Kuwahara}}, \bibinfo {author} {\bibfnamefont {T.}~\bibnamefont {Mori}}, \
  and\ \bibinfo {author} {\bibfnamefont {K.}~\bibnamefont {Saito}},\ }\href
  {\doibase 10.1016/j.aop.2016.01.012} {\bibfield  {journal} {\bibinfo
  {journal} {Annals of Physics}\ }\textbf {\bibinfo {volume} {367}},\ \bibinfo
  {pages} {96} (\bibinfo {year} {2016})}\BibitemShut {NoStop}%
\bibitem [{\citenamefont {Abanin}\ \emph
  {et~al.}(2015{\natexlab{a}})\citenamefont {Abanin}, \citenamefont {De~Roeck},
  \citenamefont {Huveneers},\ and\ \citenamefont {Ho}}]{Abanin_arXiv2015a}%
  \BibitemOpen
  \bibfield  {author} {\bibinfo {author} {\bibfnamefont {D.}~\bibnamefont
  {Abanin}}, \bibinfo {author} {\bibfnamefont {W.}~\bibnamefont {De~Roeck}},
  \bibinfo {author} {\bibfnamefont {F.}~\bibnamefont {Huveneers}}, \ and\
  \bibinfo {author} {\bibfnamefont {W.~W.}\ \bibnamefont {Ho}},\ }\href
  {http://arxiv.org/abs/1509.05386} {\bibfield  {journal} {\bibinfo  {journal}
  {arXiv:1509.05386}\ } (\bibinfo {year} {2015}{\natexlab{a}})}\BibitemShut
  {NoStop}%
\bibitem [{\citenamefont {Abanin}\ \emph
  {et~al.}(2015{\natexlab{b}})\citenamefont {Abanin}, \citenamefont
  {De~Roeck},\ and\ \citenamefont {Ho}}]{Abanin_arXiv2015b}%
  \BibitemOpen
  \bibfield  {author} {\bibinfo {author} {\bibfnamefont {D.~A.}\ \bibnamefont
  {Abanin}}, \bibinfo {author} {\bibfnamefont {W.}~\bibnamefont {De~Roeck}}, \
  and\ \bibinfo {author} {\bibfnamefont {W.~W.}\ \bibnamefont {Ho}},\ }\href
  {http://arxiv.org/abs/1510.03405} {\bibfield  {journal} {\bibinfo  {journal}
  {arXiv:1510.03405}\ } (\bibinfo {year} {2015}{\natexlab{b}})}\BibitemShut
  {NoStop}%
\bibitem [{\citenamefont {Berges}\ \emph {et~al.}(2004)\citenamefont {Berges},
  \citenamefont {Bors\'anyi},\ and\ \citenamefont {Wetterich}}]{Berges2004}%
  \BibitemOpen
  \bibfield  {author} {\bibinfo {author} {\bibfnamefont {J.}~\bibnamefont
  {Berges}}, \bibinfo {author} {\bibfnamefont {S.}~\bibnamefont {Bors\'anyi}},
  \ and\ \bibinfo {author} {\bibfnamefont {C.}~\bibnamefont {Wetterich}},\
  }\href {\doibase 10.1103/PhysRevLett.93.142002} {\bibfield  {journal}
  {\bibinfo  {journal} {Phys. Rev. Lett.}\ }\textbf {\bibinfo {volume} {93}},\
  \bibinfo {pages} {142002} (\bibinfo {year} {2004})}\BibitemShut {NoStop}%
\bibitem [{\citenamefont {Moeckel}\ and\ \citenamefont
  {Kehrein}(2008)}]{Moeckel2008}%
  \BibitemOpen
  \bibfield  {author} {\bibinfo {author} {\bibfnamefont {M.}~\bibnamefont
  {Moeckel}}\ and\ \bibinfo {author} {\bibfnamefont {S.}~\bibnamefont
  {Kehrein}},\ }\href {\doibase 10.1103/PhysRevLett.100.175702} {\bibfield
  {journal} {\bibinfo  {journal} {Phys. Rev. Lett.}\ }\textbf {\bibinfo
  {volume} {100}},\ \bibinfo {pages} {175702} (\bibinfo {year}
  {2008})}\BibitemShut {NoStop}%
\bibitem [{\citenamefont {Gring}\ \emph {et~al.}(2012)\citenamefont {Gring},
  \citenamefont {Kuhnert}, \citenamefont {Langen}, \citenamefont {Kitagawa},
  \citenamefont {Rauer}, \citenamefont {Schreitl}, \citenamefont {Mazets},
  \citenamefont {Smith}, \citenamefont {Demler},\ and\ \citenamefont
  {Schmiedmayer}}]{gring2012relaxation}%
  \BibitemOpen
  \bibfield  {author} {\bibinfo {author} {\bibfnamefont {M.}~\bibnamefont
  {Gring}}, \bibinfo {author} {\bibfnamefont {M.}~\bibnamefont {Kuhnert}},
  \bibinfo {author} {\bibfnamefont {T.}~\bibnamefont {Langen}}, \bibinfo
  {author} {\bibfnamefont {T.}~\bibnamefont {Kitagawa}}, \bibinfo {author}
  {\bibfnamefont {B.}~\bibnamefont {Rauer}}, \bibinfo {author} {\bibfnamefont
  {M.}~\bibnamefont {Schreitl}}, \bibinfo {author} {\bibfnamefont
  {I.}~\bibnamefont {Mazets}}, \bibinfo {author} {\bibfnamefont {D.~A.}\
  \bibnamefont {Smith}}, \bibinfo {author} {\bibfnamefont {E.}~\bibnamefont
  {Demler}}, \ and\ \bibinfo {author} {\bibfnamefont {J.}~\bibnamefont
  {Schmiedmayer}},\ }\href {\doibase 10.1126/science.1224953} {\bibfield
  {journal} {\bibinfo  {journal} {Science}\ }\textbf {\bibinfo {volume}
  {337}},\ \bibinfo {pages} {1318} (\bibinfo {year} {2012})}\BibitemShut
  {NoStop}%
\bibitem [{\citenamefont {Kollar}\ \emph {et~al.}(2011)\citenamefont {Kollar},
  \citenamefont {Wolf},\ and\ \citenamefont
  {Eckstein}}]{kollar2011generalized}%
  \BibitemOpen
  \bibfield  {author} {\bibinfo {author} {\bibfnamefont {M.}~\bibnamefont
  {Kollar}}, \bibinfo {author} {\bibfnamefont {F.~A.}\ \bibnamefont {Wolf}}, \
  and\ \bibinfo {author} {\bibfnamefont {M.}~\bibnamefont {Eckstein}},\ }\href
  {\doibase 10.1103/PhysRevB.84.054304} {\bibfield  {journal} {\bibinfo
  {journal} {Physical Review B}\ }\textbf {\bibinfo {volume} {84}},\ \bibinfo
  {pages} {054304} (\bibinfo {year} {2011})}\BibitemShut {NoStop}%
\bibitem [{\citenamefont {Bukov}\ \emph
  {et~al.}(2015{\natexlab{b}})\citenamefont {Bukov}, \citenamefont
  {Gopalakrishnan}, \citenamefont {Knap},\ and\ \citenamefont
  {Demler}}]{Bukov2015}%
  \BibitemOpen
  \bibfield  {author} {\bibinfo {author} {\bibfnamefont {M.}~\bibnamefont
  {Bukov}}, \bibinfo {author} {\bibfnamefont {S.}~\bibnamefont
  {Gopalakrishnan}}, \bibinfo {author} {\bibfnamefont {M.}~\bibnamefont
  {Knap}}, \ and\ \bibinfo {author} {\bibfnamefont {E.}~\bibnamefont
  {Demler}},\ }\href {\doibase 10.1103/PhysRevLett.115.205301} {\bibfield
  {journal} {\bibinfo  {journal} {Phys. Rev. Lett.}\ }\textbf {\bibinfo
  {volume} {115}},\ \bibinfo {pages} {205301} (\bibinfo {year}
  {2015}{\natexlab{b}})}\BibitemShut {NoStop}%
\bibitem [{\citenamefont {Canovi}\ \emph {et~al.}(2016)\citenamefont {Canovi},
  \citenamefont {Kollar},\ and\ \citenamefont {Eckstein}}]{Canovi2016}%
  \BibitemOpen
  \bibfield  {author} {\bibinfo {author} {\bibfnamefont {E.}~\bibnamefont
  {Canovi}}, \bibinfo {author} {\bibfnamefont {M.}~\bibnamefont {Kollar}}, \
  and\ \bibinfo {author} {\bibfnamefont {M.}~\bibnamefont {Eckstein}},\ }\href
  {\doibase 10.1103/PhysRevE.93.012130} {\bibfield  {journal} {\bibinfo
  {journal} {Phys. Rev. E}\ }\textbf {\bibinfo {volume} {93}},\ \bibinfo
  {pages} {012130} (\bibinfo {year} {2016})}\BibitemShut {NoStop}%
\bibitem [{\citenamefont {Bialynicki-Birula}\ \emph {et~al.}(1969)\citenamefont
  {Bialynicki-Birula}, \citenamefont {Mielnik},\ and\ \citenamefont
  {Pleba{\'n}ski}}]{Bialynicki-Biula1969}%
  \BibitemOpen
  \bibfield  {author} {\bibinfo {author} {\bibfnamefont {I.}~\bibnamefont
  {Bialynicki-Birula}}, \bibinfo {author} {\bibfnamefont {B.}~\bibnamefont
  {Mielnik}}, \ and\ \bibinfo {author} {\bibfnamefont {J.}~\bibnamefont
  {Pleba{\'n}ski}},\ }\href {\doibase 10.1016/0003-4916(69)90351-0} {\bibfield
  {journal} {\bibinfo  {journal} {Ann. Phys.}\ }\textbf {\bibinfo {volume}
  {51}},\ \bibinfo {pages} {187} (\bibinfo {year} {1969})}\BibitemShut
  {NoStop}%
\bibitem [{\citenamefont {Mori}\ \emph {et~al.}()\citenamefont {Mori},
  \citenamefont {Kuwahara},\ and\ \citenamefont {Saito}}]{supplement}%
  \BibitemOpen
  \bibfield  {author} {\bibinfo {author} {\bibfnamefont {T.}~\bibnamefont
  {Mori}}, \bibinfo {author} {\bibfnamefont {T.}~\bibnamefont {Kuwahara}}, \
  and\ \bibinfo {author} {\bibfnamefont {K.}~\bibnamefont {Saito}},\
  }\href@noop {} {\bibinfo  {journal} {Supplementary Material}\ }\BibitemShut
  {NoStop}%
\bibitem [{\citenamefont {Abanin}\ \emph
  {et~al.}(2015{\natexlab{c}})\citenamefont {Abanin}, \citenamefont
  {De~Roeck},\ and\ \citenamefont {Huveneers}}]{Abanin2015}%
  \BibitemOpen
\bibfield  {journal} {  }\bibfield  {author} {\bibinfo {author} {\bibfnamefont
  {D.~A.}\ \bibnamefont {Abanin}}, \bibinfo {author} {\bibfnamefont
  {W.}~\bibnamefont {De~Roeck}}, \ and\ \bibinfo {author} {\bibfnamefont
  {F.}~\bibnamefont {Huveneers}},\ }\href {\doibase
  10.1103/PhysRevLett.115.256803} {\bibfield  {journal} {\bibinfo  {journal}
  {Phys. Rev. Lett.}\ }\textbf {\bibinfo {volume} {115}},\ \bibinfo {pages}
  {256803} (\bibinfo {year} {2015}{\natexlab{c}})}\BibitemShut {NoStop}%
\bibitem [{\citenamefont {Bukov}\ \emph
  {et~al.}(2015{\natexlab{c}})\citenamefont {Bukov}, \citenamefont {Heyl},
  \citenamefont {Huse},\ and\ \citenamefont {Polkovnikov}}]{Bukov_arXiv2015}%
  \BibitemOpen
  \bibfield  {author} {\bibinfo {author} {\bibfnamefont {M.}~\bibnamefont
  {Bukov}}, \bibinfo {author} {\bibfnamefont {M.}~\bibnamefont {Heyl}},
  \bibinfo {author} {\bibfnamefont {D.~A.}\ \bibnamefont {Huse}}, \ and\
  \bibinfo {author} {\bibfnamefont {A.}~\bibnamefont {Polkovnikov}},\ }\href
  {http://arxiv.org/abs/1512.02119} {\bibfield  {journal} {\bibinfo  {journal}
  {arXiv:1512.02119}\ } (\bibinfo {year} {2015}{\natexlab{c}})}\BibitemShut
  {NoStop}%
\bibitem [{\citenamefont {Dziarmaga}(2005)}]{Dziarmaga2005}%
  \BibitemOpen
  \bibfield  {author} {\bibinfo {author} {\bibfnamefont {J.}~\bibnamefont
  {Dziarmaga}},\ }\href {\doibase 10.1103/PhysRevLett.95.245701} {\bibfield
  {journal} {\bibinfo  {journal} {Phys. Rev. Lett.}\ }\textbf {\bibinfo
  {volume} {95}},\ \bibinfo {pages} {245701} (\bibinfo {year}
  {2005})}\BibitemShut {NoStop}%
\bibitem [{Note1()}]{Note1}%
  \BibitemOpen
  \bibinfo {note} {In cold atoms in an optical lattice, $g$ is typically about
  $100$ Hz, $k=2$, and then the condition $T<1/8gk$ implies $\omega \gtrsim 10$
  kHz, which has been achieved in experiment [A. Zenesini, H. Lignier, D.
  Ciampini, O. Morsch, and E. Arimondo, \protect \href
  {http://link.aps.org/doi/10.1103/PhysRevLett.102.100403}{Phys. Rev. Lett.
  {\protect \bf 102}, 100403 (2009)}]. According to our estimation, the heating
  timescale in this case is about 1 msec, which is a typical timescale of
  cold-atom experiments.}\BibitemShut {Stop}%
\end{thebibliography}
\end{document}